\documentclass[aps,prd,superscriptaddress,floatfix]{revtex4-1}
\usepackage{color}
\usepackage{amssymb}
\usepackage[normalem]{ulem}
\usepackage{graphicx,color}
\usepackage{amsmath}
\usepackage{subfigure}
\usepackage[normalem]{ulem}
\usepackage{booktabs}
\usepackage{xcolor}
\usepackage{epsfig,graphics,color,graphicx,amsmath}

\colorlet{darkgreen}{green!50!black}
\colorlet{brightyellow}{yellow!75!red}
\colorlet{orange}{red!50!yellow}
\colorlet{darkblue}{blue!60!black}
\colorlet{darkred}{red!80!black}

\usepackage{soul}

\usepackage{mathtools}
\usepackage{mathrsfs}
\usepackage{bm}
\usepackage{relsize}
\usepackage{indentfirst}

\usepackage{xcolor}

\usepackage{amssymb,amsmath,multirow,epsfig,graphicx,color}

\usepackage{epsfig}
\usepackage{graphicx,amsmath}
\def\be{\begin{eqnarray} &&}

	\def\ee{\end{eqnarray}}

\newcommand\ba{\begin{eqnarray}}
	\newcommand\ea{\end{eqnarray}}

\newcommand{\bas}{\begin{eqnarray*}}
	\newcommand{\eas}{\end{eqnarray*}}

\newcommand{\bno}{\begin{eqnarray*}}
	\newcommand{\eno}{\end{eqnarray*}}

\def\sl
\usepackage{hyperref}
\usepackage{url}
\usepackage{hyperref}
\hypersetup{
	colorlinks=true,  
	linkcolor=blue,   
	filecolor=magenta,      
	urlcolor=blue,    
	citecolor=blue,   
} 
\begin{document}
	\vspace{-12ex}    
	\begin{flushright} 	
		{\normalsize \bf \hspace{50ex}}
	\end{flushright}
	\vspace{11ex}
	\title{Incorporating the Cosmological Constant in a Modified Uncertainty Principle}
	\author{S. Ahmadi}
	\email{samira666ahmadi@gmail.com}
	\affiliation{Department of Physics, Ayatollah Amoli Branch, Islamic Azad University, Amol, Iran}
	\author{E. Yusofi}
	\email{e.yusofi@ipm.ir(Corresponding author)}
	\affiliation{Department of Physics, Ayatollah Amoli Branch, Islamic Azad University, Amol, Iran}
	\affiliation{School of Astronomy, Institute for Research in Fundamental Sciences(IPM), P. O. Box 19395-5531,Tehran, Iran}
	\affiliation{Innovation and Management Research Center, Ayatollah Amoli Branch, Islamic Azad University, Amol, Mazandaran, Iran}
	
	\author{M. A. Ramzanpour}
	\email{ma.ramzanpour@iau.ac.ir}
	\affiliation{Department of Physics, Ayatollah Amoli Branch, Islamic Azad University, Amol, Iran}
	\affiliation{Innovation and Management Research Center, Ayatollah Amoli Branch, Islamic Azad University, Amol, Mazandaran, Iran}
	
	\date{\today}
	\begin{abstract}
	This study explores the cosmological constant problem and modified uncertainty principle within a unified framework inspired by a void-dominated scenario. In a recent paper~\cite{Yusofi:2022hgg}, voids were modeled as spherical bubbles of similar average sizes, and the surface energy on the voids' borders was calculated across various scales in a heuristic manner. We show that this results in a significant discrepancy of approximately $\mathcal{O}(+122)$ between the cosmological constant values from the minimum to the maximum radii of bubbles. Furthermore, when considering the generalized form of the uncertainty principle with both minimum and maximum lengths, i.e. $\Delta X \Delta P \geq \frac{\hbar}{2} \frac{1}{1- \beta \Delta P^2} \frac{1}{1- \alpha \Delta X^2}$, a similar order of discrepancy is observed between $\alpha_{\rm max}$ and $\alpha_{\rm min}$, indicating that $\alpha\propto\beta^{-1}\propto\Lambda\propto{\rm length}^{-2}~(m^{-2})$. As a primary outcome of this finding, we offer a novel uncertainty principle that incorporates a non-zero cosmological constant.
	\noindent \hspace{0.35cm} 
	\noindent \hspace{0.35cm} \\
	
	\end{abstract}

	\maketitle
\section{Introduction}
The modification of Heisenberg's uncertainty principle (HUP) appears to be a crucial concept in contemporary theoretical physics \cite{Maggiore:1993rv,Kempf:1993bq,Hossenfelder:2003jz,Ali:2009zq,Das:2010zf,Nozari:2008gp}. This approach is utilized to explain the limitations in position and momentum measurements, which are relevant in various scenarios. For example, a minimum length is a widely accepted prediction in different theories of quantum gravity such as string theory \cite{Amati:1988tn, Konishi:1989wk}, loop quantum gravity \cite{Garay:1994en,Rovelli:1994ge}, and black hole physics \cite{Scardigli:1999jh,Nozari:2008gp}. Additionally, a maximum momentum is evident in doubly special relativity \cite{Amelino-Camelia:2000stu}, while a maximum length naturally emerges in cosmology due to the existence of particle horizons \cite{Perivolaropoulos:2017rgq, BENSALEM2019583}.

However, quantum gravitational effects on large scales appear to be the primary influence on the formation of the cosmological perturbations to form cosmic web \cite{rovelli_2004,Chataignier:2023rkq,tHooft:1999xul,Rovelli:1994ge}. Ashtekar \textit{et al} delved into the connection between quantum gravity and the cosmic web, exploring how loop quantum gravity could give rise to the intricate network of filaments and immense voids that make up the universe's large-scale framework~\cite{Ashtekar:2016wpi,Ashtekar:2021izi}. Within the web-like structure of the cosmos, the largest entities are cosmic voids, serving as the fundamental biggest building blocks at the observable universe. Note that the average diameter of cosmic voids is typically $\simeq 100{~\rm Mpc}$. It is worth noting that given the web-like structure of the universe and its hierarchical formation~\cite{Weygaert:2011, Sheth:2003py}, the largest cosmic voids may be much larger than $100  {~\rm Mpc}$~\cite{Rudnick:2007kw}. For instance, the Boötes and KBC voids are the largest known spaces up to now, encompassing a significant portion of the visible universe's diameter~\cite{Einasto:2015vea, Cai:2022dov, Nadathur:2014tfa}.

It is plausible that the energy and pressure of these (almost) empty spaces contributes to the accelerated expansion observed in the late universe~\cite{Yusofi:2022hgg,Heydarzadeh:2022lit,Moshafi:2024guo}.
At smallest scale, the highest energetic regions of the universe could appear as a quantum foam system, made up of bubble-like cells with radii on the order of the Planck length. We will show that in a void-dominant framework, heuristic calculations reveal a scale-dependent cosmological constant, demonstrating a discrepancy of the order of $\mathcal O(10^{122})$. This order of difference can address the cosmological constant problem. The cosmological constant problem is a long-standing issue in theoretical physics that arises from the discrepancy between the predicted value of the cosmological constant in quantum field theory and the observed value in cosmology~\cite{Weinberg:2000yb, Carroll:2000fy}. 

Additionally, within the framework of the modified uncertainty principle, a similar order of discrepancy emerges for the $\alpha$ and $\beta$ coefficients present in the $
\Delta X \Delta P \geq \frac{\hbar}{2}  \frac{1}{1- \beta \Delta P^2} \frac{1}{1- \alpha \Delta X^2}~
$. As a significant result of this finding, we can conclude that $\alpha \propto \beta^{-1} \propto \Lambda$, and a new uncertainty principle may be incorporate the cosmological constant.
In Section \ref{Section II} , we review the modified uncertainty principle with minimum and maximum lengths. In Section \ref{Section III}, the core of this article, we demonstrate the incorporation of the cosmological constant into an uncertainty principle. Initially in \ref{Section III.A} , we explore the scale-dependent characteristics of the cosmological constant within the context of void-dominant cosmology.Subsequently in \ref{Section III.B}, we compare the minimum and maximum values of $\alpha$ and $\beta$  in the EUP and GUP. This comparison allows us to propose a potential correlation between the uncertainty principle's parameters $\alpha$ and $\beta$ with the cosmological constant $\Lambda$ in \ref{Section III.C}. In the final section \ref{Section IV}, we will summarize the results of this study.
	\section{Minimum and Maximum Length in Modified Uncertainty Principle}
	\label{Section II}
\subsection{HUP with minimum length (GUP)}
As mentioned in the previous section, various forms of the MUP have been proposed in the literature. The GUP implying a minimal length has the form \cite{Kempf:1994su,Bensalem:2019cxw}
\be
\Delta X \Delta P \geq \frac{\hbar}{2} (1+\beta \Delta P^2)
\label{gup1}
\ee
where $\beta=\beta_0 \frac{l_{\rm Planck}^2}{\hbar^2}$ with $\beta_{0}$ being a dimensionless parameter. This kind of GUP results in a non-zero minimal uncertainty in position measurement, represented by \be
\Delta X_{min}=\hbar\sqrt{\beta}~.
\ee
The GUP, which includes both a minimum length and a maximum momentum, has been proposed in \cite{Pedram:2011gw} in the form
\be
\Delta X \Delta P \geq \frac{\hbar}{2} \frac{1}{1-\beta \Delta P^2}
\label{gup3}
\ee
which suggests a minimal position uncertainty as
\be 
\Delta X_{min}= \frac{3\sqrt{3}}{4}\hbar\sqrt{\beta}
\label{dxmin}
\ee
 and a maximal observable momentum as \be
 P_{max} =\frac{1}{\sqrt{\beta}}~.
 \ee
\subsection{HUP with maximum length (EUP)}
	\label{sec:Section 3}
	Motivated by cosmological considerations like the particle horizon \cite{Perivolaropoulos:2017rgq,BENSALEM2019583}, which propose a natural maximum measurable length. This setup rules out a minimum position uncertainty but includes a maximum position and a minimum momentum uncertainty. An operator representation that aligns with this generalized commutation relation is provided by
	\be 
	\Delta X \Delta P \geq \frac{\hbar}{2}  \frac{1}{1- \alpha \Delta X^2}
	\label{gupexpmaxpos}
	\ee
	where $\alpha=\alpha_0 \frac{H_{\rm 0}^2}{c^2}$
	with $H_{\rm 0}$ being the Hubble constant, $c$ representing the speed of light, and $\alpha_0$ denoting a dimensionless parameter. It is important to note that this equation signifies a maximum position uncertainty given by 
	\be 
	l_{max}\equiv \Delta X_{max}= \frac{1}{\sqrt{\alpha}}~.
	\label{maxposuncert}
	\ee
	Furthermore, it implies a minimum momentum uncertainty is expressed as
	\be 
	\Delta P_{min}=\frac{3\sqrt{3}}{4} \hbar \sqrt{\alpha}~.
	\label{minmomuncert}
	\ee
	In this scenario, EUP arises from the modified commutation relation \cite{Perivolaropoulos:2017rgq},
	\be
	[\hat{X}, \hat{P}]=\frac{i\hbar}{1- \alpha \hat{X}^2}~,
	\ee
	where the position and momentum operators that satisfy this modified algebra can be represented as \cite{Perivolaropoulos:2017rgq},
	\be
	\hat{X}=\hat{x}, ~~~ \hat{P}=\frac{1}{1- \alpha \hat{x}^2}\hat{p}
	\ee
	in which the operators $\hat{x}$ and $\hat{p}$ adhere to the canonical commutation relation $[\hat{x}, \hat{p}]=i\hbar$.		
\section{Cosmological Constant in Modified Uncertainty Principle}
\label{Section III}
The cosmological constant originates from gravity, modifying Einstein's general relativity equations, and is essential in understanding the universe's dynamics and evolution. The existence of a small, non-zero cosmological constant remains a major puzzle in fundamental physics~\cite{Weinberg:2000yb, Carroll:2000fy}.  At first glance, arguments from quantum field theory might imply a cosmological constant that is possibly $10^{120}$ times larger than the one currently observed. It is widely believed that a complete theory of quantum gravity could bridge the significant gap between theory and observation~\cite{Bishop:2022cgi}. The main goal of this section is to establish a connection between the cosmological constant and the concepts of minimum (maximum) lengths within the context of void-dominated cosmology.
\subsection{Cosmological Constant in Different Lengths Scales}
\label{Section III.A}
Traditional cosmological models have tended to overlook the statics, dynamics, and evolutionary trajectories of voids, which constitute the dominant volume at both local and global scales within the universe. Voids, characterized by non-negligible energy density and evolutionary behavior, are prone to merging due to their substantial size, rendering them key candidates for influencing cosmic scales. Consequently, these large-scale inhomogeneities likely play a crucial role in shaping the dynamics of the universe and influencing the evolution of cosmic parameters.
On the cosmic scales, the universe appears void-dominated, with clusters, filaments, and matter nodes clustering in over-dense regions among these voids. Research suggests that voids serve as counterparts to massive clusters, and their coexistence, continuous clustering of superclusters, and merger of supervoids significantly contribute to the universe's structural evolution. 

\begin{table}
	\caption{Surface tension (energy) $\sigma_{\rm i}$ and cosmological constant $\Lambda_{\rm i}$, for the clustered cosmic objects including superclusters (Sc), clusters (C), and galaxies \cite{Yusofi:2022hgg}. The values in the table clearly indicates that both surface energy and the relevant cosmological constant increase inversely with the diminishing size of the objects.}
	\label{table:I}       
	\begin{center}
		\begin{tabular}{lllll}
			
			\noalign{\smallskip}\hline
			\noalign{\smallskip}\hline
			i. Clustered &\quad $M_{\rm i}$  &\quad $R_{\rm i}$  &\quad \quad$\sigma_{\rm i}$&\quad \quad$\Lambda_{\rm i}$ \\
			~~Objects &($10^{47}{\rm kg}$) &($10^{24}{\rm m}$) & ($10^{15}{\rm J.m^{-2}}$)& ($10^{-52}{\rm m^{-2}}$) \\
			\noalign{\smallskip}\hline
			\noalign{\smallskip}\hline
			1. Corona Sc&\quad $0.20$ &\quad $1.50$ &\quad $0.25$ &\quad $0.6645$ \\			2. Virgo Sc&\quad $0.03$ &\quad $0.50$ &\quad $0.34$ &\quad $0.8970$ \\
			3. Laniakea Sc &\quad $1.00$ &\quad $2.40$ &\quad $0.50$ &\quad $1.2979$\\
			4. Caelum Sc&\quad $4.00$ &\quad $ 4.30$ &\quad $0.62$ &\quad $1.6172$\\
			\noalign{\smallskip}\hline\noalign{\smallskip}
			5. Virgo C &\quad $0.002$ &\quad $0.045$ &\quad $2.83$ &\quad $7.3833$ \\
			6. Coma C &\quad $0.012$ &\quad $ 0.10$ &\quad $3.44$ &\quad $8.9707$ \\
			7. Galaxies G &\quad $0.0004$ &\quad $ 0.016$ &\quad $4.48$ &\quad $11.6806$ \\
			\noalign{\smallskip}\hline\noalign{\smallskip}
			8. Milky Way &\quad $0.00002$ &\quad $ 0.00087$ &\quad $75.74$ &\quad $1975.3200$\\
			9. Andromeda &\quad $0.00003$ &\quad $ 0.001$ &\quad $85.98$ &\quad $2242.6800$\\
			10. UGC02885 &\quad $0.000004$ &\quad $ 0.00036$ &\quad $88.46$ &\quad $2307.2800$\\
			\noalign{\smallskip}\hline
		\end{tabular}
	\end{center}
\end{table}
\begin{table}
	\caption{The maximum and minimum surface tension $ \sigma_{\rm i} $, and cosmological constant $ \Lambda_{\rm i}$ for a bubbly cosmos in the possible smallest and the largest scales.}
	\label{table:II}
	\begin{center}
		\begin{tabular}
			{lllll}
			\hline
			\hline
			Length Scale &\qquad$M_{\rm i}({\rm kg})$&\qquad$R_{\rm i}({\rm m})$&\quad $\sigma_{\rm i}({\rm j.m^{-2}})$& $\qquad\Lambda_{\rm i}({\rm m^{-2}})$  \\
			\hline
			\hline
			Minimum (Planck)&\qquad$2.18\times10^{-8}$&\quad $1.62$$\times 10^{-35}$ &\quad$2.38\times10^{78}$ &\quad$6.08\times 10^{70}$\\
			\noalign{\smallskip}\hline
			Maximum (Caelum Sc)&\qquad$4.00\times10^{47}$&\quad $4.30$$\times 10^{24}$ &\quad$0.62\times10^{15}$ &\quad$1.62\times 10^{-52}$\\
			\noalign{\smallskip}\hline
		\end{tabular}
	\end{center}
\end{table}

In void-dominated cosmology, the cosmological constant has variable values on different scales~\cite{Xu_2010,Carroll:2000fy,Peebles:1987ek}. Since the sizes of voids are diverse, we must obtain the largest values for surface tension and cosmological constant in the smallest scales and vice versa (see Table \ref{table:I}). We assume voids (bubbles) are encircled by matter clusters with a shell-like distribution on the border of voids~\cite{Einasto:2015vea}. Considering the shells as the ideal separating surface between the under-density voids and the over-density clusters, we can calculate the resulting surface tension (energy) through dimensional calculation ~\cite{Yusofi:2022hgg}. By using Friedman's equations, it is obtained the relationship between the surface energy and the cosmological constant~\cite{Yusofi:2022hgg}. We list in Tables \ref{table:I} and \ref{table:II} the surface energy $\sigma_{\rm i}$ and the corresponding cosmological constant $\Lambda_{\rm i}$ for each bubble, ranging from the largest to the smallest in our simulated universe, in order to potentially address the cosmological constant problem.\\
To illustrate this more precisely, we consider the initial web-like structure as a collection of very small Planck-sized voids connected to each other in the foam-shape. By utilizing the following relation~\cite{Yusofi:2022hgg},
\begin{equation}
	\label{Lami}
	\Lambda_{\rm i} =\frac{8\pi{G}}{{w{c^4}}}\frac{2\sigma_{\rm i}}{\bar r_{\rm {v}}}.
\end{equation}
we calculate the cosmological constant for a hypothetical single bubble with the Planck radius in Table \ref{table:II} in the order of $\mathcal{O} (10^{+70}$). The minimum of cosmological constant in the latest observational data is reported as \cite{Planck:2018vyg},
\begin{equation}
	\label{hob8}
	\Lambda_{\rm {obs}}= 1.1056 \times 10^{-52} {\rm {m^{-2}}}.
\end{equation}
In our model, for the largest object such as Laniakia supercluster (please refer the row 3 in Table \ref{table:I}), the cosmological constant of the model is estimated as \cite{Yusofi:2022hgg},
\begin{equation}
	\label{hob9}
	\Lambda_3 =  1.2979 \times  10^{-52} {\rm {m^{-2}}}.
\end{equation}
In Table \ref{table:I}, we observe that the cosmological constant and mass density for each cosmic void are equivalent to the values for the entire universe and are very close to it. Thus, given the values obtained for the $\rho_{\rm i}$, $\sigma_{\rm i}$ and $\Lambda_{\rm i}$, it seems that a cosmic void can be a good indicator of the global behavior of the universe. Therefore, the decrease in surface tension observed in expanding cosmic voids may be considered a potential source of the negative pressure that contributes to the acceleration of the universe \cite{Yusofi:2022hgg}.
Our model can now address the discrepancy in cosmological constant values at the Planck scale with the smallest radius (minimum possible length) and the present universe (see Table \ref{table:II}), which related to the superclusters/voids (even super-Hubble expanding bubble~\cite{Sheth:2003py}) with the largest radius (maximum possible length) and is of the following order of magnitude,
\begin{equation}
	\label{hob9}
	\frac{\Lambda_{\rm Planck}}{\Lambda_{\rm Obs}}=\frac{\Lambda_{\rm max}}{\Lambda_{\rm min}} \sim 10^{+122}.	
\end{equation}
Here, $\Lambda_{\rm max}=\Lambda_{\rm UV}$ could be seen as a high-energy extension of the traditional cosmological constant, resulting in $\Lambda_{\rm min}=\Lambda_{\rm IR}$ in the low-energy range~\cite{Shalyt-Margolin:2008rmj,Jejjala:2006jf,Padmanabhan:2006ag}. 
According to the discussions and results of the previous sections, in this section, we examine the potential of modified uncertainty principle (MUP) which incorporate the cosmological constant.
\subsection{MUP with Minimum and Maximum Position (Momentum)}
\label{Section III.B}
A natural generalization of (\ref{gup1}) corresponds to the existence of minimum position and minimum momentum uncertainty is obtained by,~\cite{Perivolaropoulos:2017rgq},
\begin{equation}
[X,P]=\frac{\hbar}{2}(1+\alpha X^2+\beta P^2).
\end{equation} 
There are multiple ways to determine the operator representation related to  this the commutation relation . The position and momentum operators satisfying this commutation can be defined by utilizing the operators $x$ and $p$, which follow the conventional commutation relation $[x, p] = i\hbar$. As an illustration, one may consider the following,
\begin{equation}
	X=x, \;\;\;  \mbox{and} \;\; P=p(1+\alpha x^2+\beta p^2).
\end{equation}

Based on this definition, the more general form of (\ref{gup1}) and (\ref{gupexpmaxpos}) includes explicit maximum and minimum values for both position and momentum uncertainties and can be written as follows,
\be
\label{GEUP}
\Delta X \Delta P \geq \frac{\hbar}{2}  \frac{1}{1- \beta \Delta P^2} \frac{1}{1- \alpha \Delta X^2}~.
\ee
Here $\Delta X$ and $\Delta P$ represent position and momentum uncertainties, while $\alpha$ and $\beta$ are the parameters. 

Motivated by the existence of the Planck length and the largest super voids (Hubble bubble or cosmic horizon) as the scales which set natural minimum and maximum possible lengths, we now turn our attention to relation (\ref{GEUP}) by considering the following relationship among $\Delta X$, $\alpha$ and $\beta$ (please refer to section III of \cite{Perivolaropoulos:2017rgq}),
$$
\Delta X\propto \frac{1}{\sqrt{\alpha}}\propto\sqrt{\beta}~.
$$
In the proposed bubbly model of the universe (as per Tables \ref{table:I} and \ref{table:II}), the size of spherical objects can vary from the smallest to the largest radius. As a result, the maximum position uncertainty can be expressed as,
\be 
\label{key20}
l_{max}\equiv \Delta X_{max}\propto \frac{1}{\sqrt{\alpha_{min}}}\propto\sqrt{\beta_{max}}\simeq 10^{+26}{\rm m}~,
\label{maxpos}
\ee
and the minimum position uncertainty can be expressed as,
\be 
\label{key21}
l_{min}\equiv \Delta X_{min}\propto \frac{1}{\sqrt{\alpha_{max}}}\propto\sqrt{\beta_{min}}\simeq 10^{-35}{\rm m}~.
\label{minpos}
\ee
By combining equations (\ref{key20}) and (\ref{key21}), we can obtain from Table \ref{table:II},
\begin{equation}
	\label{hob19}
	\frac{\alpha_{\rm max}}{\alpha_{\rm min}}\sim\left(\frac{l_{\rm min}}{l_{\rm max}}\right)^{-2}\sim \left(\frac{10^{-35}}{10^{+26}}\right)^{-2} = 10^{+122}.
\end{equation}
\subsection{Modified Uncertainty Principle with Cosmological Constant}
\label{Section III.C}
Based on the above discussions, particularly relations (\ref{hob9}), (\ref{maxpos}), (\ref{key21}) and (\ref{hob19}), we can conclude that the parameter $\alpha$ is proportional to the cosmological constant $\Lambda$, $\alpha \propto\beta^{-1}\propto \Lambda$. Dimensional analysis, as suggested in previous works~\cite{Bolen:2004sq, Bambi:2007ty}, implies that the primary correction's structure is proportional to $\Lambda \sim {l_{\Lambda}}^{-2}\sim(\frac{c}{H_0})^{-2}$~\cite{Padmanabhan:2006ag}. So we can get the following relations,
\be 
\label{key200}
\Delta X_{max}\propto \frac{1}{\sqrt{\Lambda_{min}}}~,
\label{maxpos2}
\ee
and the minimum position uncertainty can be expressed as,
\be 
\label{key211}
\Delta X_{min}\propto \frac{1}{\sqrt{\Lambda_{max}}}~.
\label{minpos2}
\ee
In another hand, considering the merger of voids and their expansion, voids can evolve into genuine super-Hubble bubbles~\cite{Sheth:2003py},  which can have a radius in the size of the horizon i.e. $l_{\rm horizon}\equiv l_{\Lambda} = \sqrt{\frac{3}{\Lambda}}\propto \frac{1}{\sqrt{\Lambda}}$ as a (anti-) de Sitter horizon~\cite{Luminet:2013sya,Bambi:2007ty}. Also, It is also interesting to note that both $\alpha$ and $\Lambda$ are of the same dimension, $m^{-2}$.
As a final result of this discussion, we can propose \textit{a new extension of uncertainty principle including the cosmological constant} as follows:
\be
\label{GEUPnew}
\Delta X \Delta P \geq \frac{\hbar}{2}  \frac{1}{1- \beta_{0} \Lambda^{-1} \Delta P^2} \frac{1}{1- \alpha_{0} \Lambda \Delta X^2}~.
\ee

Here $\alpha\equiv\alpha_{0} \Lambda$, and $\beta\equiv\beta_{0} \Lambda^{-1}$. In the final relation (\ref{GEUPnew}), $\Lambda$ have approximate values, so we require non-zero  $\alpha_{0}$ and $\beta_{0}$ as the dimensionless coefficients. Our new extension of the uncertainty principle (\ref{GEUPnew}) is directly linked to the non-zero cosmological constant, becoming an essential element for a more comprehensive understanding of quantum spacetime in both small and large scales of the cosmos. The parameter $\alpha$ in the proposed MUP is not meaningless but rather is proportional to the energy of the vacuum (quasi-empty or void) space. In our view, the cosmological constant varies at different scales with the highest value of momentum ($\Lambda_{\rm max}$) at the smallest scale ($\Delta X_{\rm min}$) and vice versa. If our hypothesis (\ref{GEUPnew}) is true, this implies that Heisenberg's uncertainty principle is never valid, and its gravitational impacts must be considered since $\Lambda$ are non-zero at all levels. 


	\section{Conclusions}
	\label{Section IV}
In this research, representing voids as nearly identical spherical bubbles at the largest distances in the current universe and the smallest distances at the Planck scale led to two interesting findings:

\begin{enumerate}
	\item
	Estimating the cosmological constant heuristically for a possible largest and smallest bubbles (see Table \ref{table:II}) in a void-dominated model showed that the difference between the highest and lowest cosmological constants is approximately $\mathcal{O}(122)$.
	\item
	Taking into account the uncertainty principle for minimum and maximum length scales, a similar disparity was noted for $\alpha_{\text{max}}$ and $\alpha_{\text{min}}$, suggesting that $\alpha\propto\Lambda\propto {\rm length}^{-2}~(m^{-2})$.
\end{enumerate}
This biggest of discrepancy, considered a notable challenge in theoretical physics, is perceived as scale-dependent and inherently intrinsic in our proposition. A key consequence of this revelation was the formulation of a new uncertainty principle that incorporates the cosmological constant.

Although we don't claim to resolve the cosmological constant issue or introduce a new version of the uncertainty principle, we have determined that the presence of a non-zero cosmological constant is essential at all levels, and its value is contingent on the selected regime. In our view, the significant contrast between the highest and lowest values of the cosmological constant is not a concern. Our proposal suggests that the difference in the values of cosmological constant is inherent and should be considered natural. 
It is essential to acknowledge that the findings presented in this paper are approximate and heuristic. Nevertheless, given the extremely small or large orders of magnitude of the values under investigation, these simplifications and approximations closely reflect reality.
Furthermore, by introducing an extended version of the uncertainty principle, we have demonstrated that a non-zero cosmological constant is essential for establishing the minimum and maximum length in physics. From our perspective, this non-zero cosmological constant could be attributed to the existence of energy density and pressure in near-empty space at any scales, and these cases are the topics we will address in future projects.\\
	\section*{Declaration of Competing Interest}
	The authors declare that they have no known competing financial interests or personal relationships that could have appeared to influence the work reported in this paper.
	\section*{Acknowledgements}
	This work has been supported by the Islamic Azad University, Ayatollah Amoli Branch, Amol, Iran.\\

	\bibliography{MUP2-CC-Samira.bib}

\end{document}